\documentclass[a4paper,12pt]{article}

\usepackage{amsmath,amssymb,amsfonts,graphicx}

\usepackage{epsfig}
\usepackage{color}
\usepackage{comment}
\usepackage{ulem}
\usepackage{mathdots}
\usepackage{graphics}
\usepackage{epstopdf}
\usepackage{mathdots}
\usepackage{epsfig}
\usepackage{verbatim}
\usepackage{color}
\usepackage{array}
\usepackage{url}
\usepackage{textcomp}
\usepackage{algorithm,algpseudocode}

\usepackage{cellspace}
\newcolumntype{P}[1]{>{\centering\arraybackslash}p{#1}}

\def\sq{ { {\color{blue} \hfill\rule{1.5mm}{1.5mm} } } }

\newtheorem{definition}{Definition}[section]
\newtheorem{proposition}{Proposition}[section]
\newtheorem{corollary}{Corollary}[section]
\newtheorem{remark}{Remark}[section]

\newcommand{\bA}{{\bf A}}
\newcommand{\bB}{{\bf B}}
\newcommand{\bC}{{\bf C}}
\newcommand{\bD}{{\bf D}}

\newcommand{\bY}{{\bf Y}}

\newcommand{\bK}{{\bf K}}

\newcommand{\bN}{{\bf N}}
\newcommand{\bI}{{\bf I}}

\newcommand{\bP}{{\bf P}}

\newcommand{\bQ}{{\bf Q}}

\newcommand{\bZ}{{\bf Z}}
\newcommand{\bX}{{\bf X}}
\newcommand{\bx}{{\bf x}}
\newcommand{\by}{{\bf y}}
\newcommand{\bu}{{\bf u}}

\newcommand{\bU}{{\bf U}}

\newcommand{\bg}{{\bf g}}

\newcommand{\bh}{{\bf h}}

\newcommand{\bff}{{\bf f}}
\newcommand{\bfz}{{\mathbf 0}}

\newcommand{\cH}{ {\cal H} }

\newcommand{\cP}{ {\cal P} }
\newcommand{\cQ}{ {\cal Q} }

\newcommand{\Si}{ \boldsymbol{\Sigma} }

\newcommand{\tr}{ {\rm trace} }

\title{\LARGE \bf
	On the $\mathcal{H}_2$ norm and iterative model order reduction of linear switched systems 
}

\author{Ion Victor Gosea \footnotemark[2] and Athanasios C. Antoulas \footnotemark[3] % <-this % stops a space
}

\begin{document}
	
	\maketitle
	
	\renewcommand{\thefootnote}{\fnsymbol{footnote}}

		\footnotetext[2]{Data-Driven System Reduction and Identification Group, Max Planck Institute for Dynamics of Complex Technical Systems, Magdeburg, Germany {\tt gosea@mpi-magdeburg.mpg.de}}%
\footnotetext[3]{Department of Electrical and Computer Engineering, Rice University, Houston, USA and MPI Magdeburg, Germany and Baylor College of Medicine, Houston, USA  {\tt aca@rice.edu}}

\begin{abstract}
  A new definition of the $\mathcal{H}_2$ norm for linear switched systems is introduced. It is based on appropriately defined time-domain kernels, or equivalently, on infinite controllability and observability Gramian matrices. Furthermore, an extension of the iterative rational Krylov algorithm to the class of linear switched systems is proposed.
\end{abstract}

%%%%%%%%%%%%%%%%%%%%%%%%%%%%%%%%%%%%%%%%%%%%%%%%%%%%%%%%%%%%%%%%%%%%%%%%%%%%%%%%%%%%%%%%%

\section{Introduction}

Hybrid systems are a class of nonlinear systems which result from the interaction of continuous time dynamical subsystems with discrete events.
% These systems are described by both discrete and continuous states, inputs and outputs. 
Switched systems constitute a subclass of hybrid systems, in the sense that the discrete dynamics is simplified. Hence, any discrete state transition is allowed and the set of discrete events coincides with the set of discrete states.

In this work we analyze continuous-time linear switched systems (LSS) which have reset maps (known also as coupling/switching matrices). The latter term refers to matrices that scale the continuous state at the switching times. For a detailed characterization of LSS, we refer the reader to the books \cite{c11} and \cite{c15}.

Model order reduction (MOR) is a tool for approximating large and complex models characterizing time-dependent processes by much  smaller and simpler models that are still able capture the dominant characteristics of the original process. Such reduced order models (ROM) could be used as efficient surrogates for the original model, replacing it as a component in larger simulations. For details on different MOR techniques, we refer the reader to the book \cite{c1} and to the surveys \cite{c6} and \cite{c10}.

One of the most prolific classes of MOR methods is represented by the projection-based techniques (see \cite{c6}), including the special cases of  balancing and interpolation. For the first subclass,
one chooses projectors so that the system is transformed into a balanced realization (in which states that are hard to control or to observe can be easily removed). To construct such transformation, one is required to find infinite Gramian matrices, for example by solving Lyapunov equations. Gramians play an important role in the analysis of linear dynamical systems (see \cite{c1}). For the second subclass, the original model is projected onto appropriate Krylov subspaces.
More specifically, the projectors are chosen in such a way that the transfer function of the reduced model interpolates those of the full model at selected interpolation points (see \cite{c2}).

The iterative rational Krylov algorithm, or in short IRKA, was introduced in \cite{c10}. It was proven to be a very effective iterative procedure, which, upon convergence, yields a locally $\cH_2$ optimal reduced system. By means of repeatedly applying a two-sided projection constructed in the Petrov-Galerkin framework (\cite{c1}), the algorithm  enforces interpolation-based optimality conditions (Theorem 3.4 in \cite{c10}).

Model order reduction methods have also been applied to the class of LSS in recent years. We mention the one in \cite{c3}, which involves time-domain matching of generalized Markov parameters and also the balanced truncation methods proposed in \cite{c14} and \cite{c11}.

Next, we make a short inventory of $\cH_2$-type methods for bilinear systems (see \cite{c7} for an overview of such systems). The motivation behind this is to familiarize the reader with some extensions of IRKA proposed in recent years as well as to mention the inspiration that triggered the ideas behind this current work. In many aspects, bilinear systems and LSS show similarities, some of which are presented in Section 5. 

The first attempt of extending the $\cH_2$ optimal approximation framework from the class of linear systems to that of bilinear systems, was made in \cite{c16}. Here, the so called Gramian-based optimality conditions were extended to the bilinear case (see Section 5 in \cite{c16}). Afterwards, in \cite{c5}, the Bilinear Iterative Rational Krylov Algorithm (B-IRKA) was introduced as an iterative,
projection-based approach which extended IRKA to the bilinear case. In \cite{c9}, a new framework which enforces multipoint interpolation of the underlying Volterra series is introduced, as well as a new derivation for the $\cH_2$ norm of bilinear system.

In this paper, we focus on introducing the $\cH_2$ norm as well as extending IRKA to the class of linear switched systems. In Section 2, a brief overview on linear switched systems is presented, including the definition of generalized time-domain kernels. Afterwards, Section 3 contains the definition of infinite Gramians for the case in which the two subsystems are decoupled (no switching occurs) as well as the extension to the case when the dynamics is coupled (by taking into account all possible switching scenarios). In Section 4 the $\cH_2$ norm is extended to the class of LSS. Then, the proposed iterative algorithm is introduced in Section 5, which also contains a numerical example. Finally, Section 6 includes the conclusions and some possible research directions.

\section{Linear switched systems} \label{sec:switched}

A switched system is a dynamical system that consists of
a finite number of subsystems and a logical rule that
orchestrates switching between these subsystems.
These subsystems  or discrete modes are usually described by a collection of  differential or difference equations. The discrete events interacting with the subsystems are governed by a piecewise continuous function, i.e. the switching signal.

For simplicity of the exposition, we only consider the case of an LSS that switches between two modes. This situation is encountered in most of the numerical examples in the literature we came across. Nevertheless, the theoretical concepts presented in this work, can be generalized to any number of modes denoted with M.

\begin{definition}
	Let $\Si$ be a continuous time linear switched system (LSS), defined as
	\begin{equation}\label{LSS_def0}
	\Si: \begin{cases}
	\dot{\bx}(t) = \bA_{\sigma(t)} \bx(t) + \bB_{\sigma(t)}\bu(t), \ \ \bx(t) = \bx_0, \\
	\by(t) = \bC_{\sigma(t)} \bx(t),
	\end{cases}
	\end{equation}
	where $\Omega = \{1,2\}$, is the set of discrete modes, $\sigma(t)$ is the switching signal, $\bu$ is  the input, $\bx$ is the state, and $\by$ is the output.
	
\end{definition}

The system matrices  $\bA_q \in \mathbb{R}^{n_q \times n_q}, \ \bB_q \in \mathbb{R}^{n_q \times m_q}, \ \bC_q \in \mathbb{R}^{p_q \times n_q}$, where $q \in \Omega$, correspond to mode $q \in \Omega$, and $\bx_0 \in \mathbb{R}^{n_{q_1}}$ is the initial state. Furthermore, the transition from one mode to another is made via the so called switching or coupling matrices $\bK_{q,s} \in \mathbb{R}^{n_{s} \times n_{q}}$ where  $q,s \in \Omega$ ($q \neq s$). The case for which the coupling is made between identical modes is excluded.

%Hence, when $q_1 = q_2 = q$, consider that the coupling matrices are identity matrices, i.e. $\bK_{q,q} = \bI_{n_q}$.

The notation $\Si = (n_1,n_2, \{(\bA_q, \bB_q, \bC_q)| q \in \Omega\}$, $\{ \bK_{1,2}, \bK_{2,1} \}, \bx_0)$ is used as a short-hand representation for the LSS described by the equations in (\ref{LSS_def0}). 

%The vector $\bn = \left[\begin{array}{cc}
%n_1 & n_2 \end{array} \right]$ is the dimension (order) of $\Si$. 

The restriction of the switching signal $\sigma(t)$ to a finite interval of time $[0, T]$ can be interpreted as finite sequence of elements of $\Omega \times \mathbb{R}_{+}$ of the form:
$$
\nu(\sigma) = (q_1,t_1) (q_2,t_2) \ldots (q_k,t_k),
$$
where $q_1, \ldots , q_k \in \Omega$ and $0<t_1 < \cdots < t_k  \in \mathbb{R}_{+}$, \ $\sum_{j=1}^k t_i = T$. For all time instances $ t \in [0, T]$, we have:
\begin{equation}\label{switch_signal}
\sigma(t) = \begin{cases}  q_1 \ \  \text{if} \ \ t \in [0,T_1], \\
q_i \ \  \text{if} \ \ t \in (T_{i-1},T_i], \ i > 2, \end{cases},
\end{equation}
where the switching times are defined as $T_i := \sum_{j=1}^i t_j$, \ with $T_0 :=0, \  T_k := T$. The values $t_j \in \mathbb{R}_{+}$ are called dwelling times.
% the additional section is marked with orange

The linear system that is activated in mode $q \in \{1,2\}$, is denoted with $\Si_{q}$ and it is a subsystem of $\Si$, described by
\begin{equation}\label{LSS_def2}
\Si_q: \begin{cases}
\dot{\bx}(t) = \bA_{q}\bx(t) + \bB_{q}\bu(t), \ \   \\
\by(t) = \bC_{q} \bx(t). \ \ 
\end{cases}
\end{equation}
The dimension (order) of the subsystem $\Si_q$ is given by the scalar $n_q$. Denote with $PC(\mathbb{R}_{+},\mathbb{R}^n)$, $P_c(\mathbb{R}_{+},\mathbb{R}^n)$, the set of all piecewise-continuous, and piecewise-constant functions, respectively.
\begin{definition}
	A tuple $(\bx,\bu,\sigma,\by)$, where $\bx:\mathbb{R}_{+} \rightarrow \bigcup_{i=1}^{2} \mathbb{R}^{n_i}$, $\bu \in PC(\mathcal{R}_{+},\bigcup_{i=1}^{2} \mathbb{R}^{m_i}),$ $ \sigma \in P_c(\mathbb{R}_{+},\Omega), \by \in PC(\mathbb{R}_{+},\bigcup_{i=1}^{2} \mathbb{R}^{p_i})$ is called a solution, if the following conditions simultaneously hold:
	%Assume that $0=t_1 < t_2 < \cdots < t_k < $ are such that $\sigma(t)=q_i$ for all $t \in [t_i,t_{i+1})$, $i=1,2,\cdots $.  
	
	\begin{enumerate}
		\item The restriction of $\bx(t)$  to $(T_{i-1},T_{i}]$ is differentiable, and satisfies $ \dot \bx (t) =\bA_{q_i}\bx(t)+\bB \bu(t)$.
		\item  When switching from mode $q_{i}$ to mode $q_{i+1}$ at time $T_i$, $ \displaystyle \lim_{t \searrow T_i} \bx(t) = \bK_{q_i,q_{i+1}}  \bx(T_i)$ holds.
		% $\bx(t_{i+1})=\bK_{q_i,q_{i+1}} \lim_{ t \rightarrow t_{{i+1}^{-}}} \bx(t)$ holds.
		\item  For all $t \in \mathbb{R}$, it follows that $\by(t)=\bC_{\sigma(t)} \bx(t)$.
	\end{enumerate}
\end{definition}

The coupling matrices $\bK_{q_{i},q_{i+1}}$ allow having different dimensions for the subsystems active in different modes. If the coupling matrices are not explicitly given, it is considered that they are identity matrices.

The input-output behavior of an LSS system can be described in time domain using the mapping $\bff(\bu,\sigma)$. This map can be written in \textit{generalized kernel representation}, as suggested in \cite{c13}, using the unique family of analytic functions: $\bg_{q_1,\ldots,q_k}: \mathbb{R}_{+}^k \rightarrow \mathbb{R}^{p_{q_1}}$ and $\bh_{q_1,\ldots,q_k}: \mathbb{R}_{+}^k \rightarrow \mathbb{R}^{p_{q_1} \times m_{q_k}}$ with $q_1,\ldots,q_k \in \Omega, \ k \geqslant 1$ such that for all pairs $(\bu,\sigma)$ and for $T = t_1+t_2+\cdots+t_k$ we can write:
\small
\begin{align}
\bff(\bu, & \sigma)(t)  = \bg_{q_1,\ldots,q_k}(t_1,\ldots,t_k)+ \nonumber \\ &  \displaystyle \sum_{i=1}^{k} \int_{0}^{t_i} \bh_{q_i,\ldots,q_k}(t_i-\tau,t_{i+1},\ldots,t_k) \bu(\tau+T_{i-1}) d\tau,
\end{align}
\normalsize
The level k functions $\bg, \bh$ are defined for $k \geqslant 1$, as: 
\small
\begin{align}\label{init_state}
\bg_{q_k,\ldots,q_1}(t_k, & \ldots,t_1)  = \nonumber \\  & \bC_{q_k}e^{\bA_{q_k}t_k} \bK_{q_{k-2},q_{k-1}} \cdots \bK_{q_{1},q_{2}} e^{\bA_{q_1}t_1} \bx_0,
\end{align}
\vspace{-7mm}
\begin{align}\label{kern}
\bh_{q_k,\ldots,q_1}(t_k, & \ldots,t_1) =  \nonumber \\  & \bC_{q_k}e^{\bA_{q_k}t_k} \bK_{q_{k-1},q_{k}} \cdots \bK_{q_{1},q_{2}} e^{\bA_{q_1}t_1}\bB_{q_1}.
\end{align}
\normalsize
In this work we consider systems with zero initial conditions ($\bx_0 = \bfz$). When this assumption does not hold, one can incorporate $\bx_0$ in, by enlarging the $\bB$ matrix corresponding to the first active mode $q_1$, i.e. as $[\bB_{q_1} \ \bx_0]$.

The generalized kernels in (\ref{kern}) will be used in defining the $\cH_2$ norm for LSS. For the switching sequences of length 2, i.e. (1,2) and (2,1), write the level 2 functions in (\ref{kern}), as:
$$
\text{Level 2}: \begin{cases} \bh_{1,2}(t_1,t_2) = \bC_1  e^{\bA_{1} t_1} \bK_{2,1} e^{\bA_{2}t_2} \bB_2, \\ \bh_{2,1}(t_1,t_2) = \bC_2  e^{\bA_{2} t_1} \bK_{1,2} e^{\bA_{1}t_2} \bB_1. \end{cases}
$$ 

\section{Infinite Gramian matrices}

Let $\Si$ be an LSS with two modes, as introduced in (\ref{LSS_def0}). We assume both $\bA_1$ and $\bA_2$ matrices have eigenvalues with negative real part, i.e. $\text{Re}(\lambda_i (\bA_k)) <0, \ k \in \{1,2\}$. 

The stability of the subsystems $\Si_k$ is essential for the existence of the infinite Gramains that will be introduced in this section.

\subsection{The case with no switching}

Start with defining the linear Gramians for the simplified case when no switching occurs. The LSS operates only in mode $q \in \Omega$, where q can be either 1 or 2.

First introduce the controllability Gramians denoted with  $\cP_{q}^{(1)}$, corresponding to mode $q \in \{1,2\}$, that are defined as
\begin{equation}\label{def_lin_gram_con}
\cP_{q}^{(1)} = \int_0^\infty \bg^c_q(t) \big{(}  \bg^c_q(t) \big{)}^T dt = \int_0^\infty e^{\bA_q t} \bB_q \bB_q^T  e^{\bA_q^T t} dt,
\end{equation}
where $\bg^c_q(t) = e^{\bA_qt} \bB_q, t\geqslant 0$. It is a well known result that $\cP_{q}^{(1)}$ satisfies the following Lyapunov equation:
\begin{align}\label{lin_gram_con}
\bA_q \cP_q^{(1)} + \cP_q^{(1)} \bA_q^T + \bB_q \bB_q^T = \bfz.
\end{align}
\noindent
Denote with $\cQ_{q}$ (or $\cQ_{q}^{(1)}$) the linear observability Gramian  $\cQ_{q}^{(1)}$ corresponding to mode $q \in \Omega$, which can be written as
\begin{equation}\label{def_lin_gram_obs}
\cQ_{q}^{(1)} = \int_0^\infty \big{(} \bg^o_q(t) \big{)}^T \bg^o_q(t)   dt = \int_0^\infty e^{\bA_q^T t} \bC_q^T \bC_q  e^{\bA_q t} dt.
\end{equation}
where $\bg^o_q(t) =  \bC_q e^{\bA_q t}, t\geqslant 0$. Again, it is known that $\cQ_{q}$ satisfies the following Lyapunov equation:
\begin{align}\label{lin_gram_obs}
\bA_q^T \cQ_q^{(1)} + \cQ_q^{(1)} \bA_q + \bC_q^T \bC_q = \bfz.
\end{align}

\subsection{The LSS case - constraint switching}

\subsubsection{Controllability Gramians}

Let $q_1 \in \{1,2\}$ be the starting operating mode. Introduce the level $k$ energy functional denoted with $\bg^c_{q_1,q_2,\ldots,q_k}(t_1,t_2,\ldots,t_k) : \mathbb{R}^k \rightarrow \mathbb{R}^{m_{q_k}}$, corresponding to a switching sequence $(q_1,\ldots,q_k) \in \Omega^k$ that starts in mode $q_1$:
\small
\begin{align}\label{grqdef}
\hspace{-3mm} \bg^c_{q_1,\ldots,q_k}(t_1,\ldots,t_k) = e^{\bA_{q_1}t_1} \bK_{q_{2},q_{1}} \cdots \bK_{q_{k},q_{k-1}} e^{\bA_{q_k}t_k} \bB_{q_k}. \hspace{-2.6mm}
\end{align}
\normalsize
In general, compute the level $k$ infinite controllability Gramian corresponding to mode $q_1 \in \{1,2\}$  by calculating the inner product of the energy functional associated to the length $k$ switching sequence $(q_1,q_2,\ldots,q_k)$ with itself, as
\small
\begin{align}\label{Pkdef}
\cP_{q_1}^{(k)} &= \int_{0}^\infty \cdots \int_{0}^\infty   \bg^c_{q_1,q_2,\ldots,q_k}(t_1,t_2,\ldots,t_k) \nonumber \\ & \big{(}\bg^c_{q_1,q_2,\ldots,q_k}(t_1,t_2,\ldots,t_k) \big{)}^T dt_1 dt_2 \ldots dt_k.
\end{align}
\normalsize
\noindent
By making use of the recurrence relation
\begin{align*}
\bg^c_{q_1,q_2,\ldots,q_k} & (t_1,t_2,\ldots,t_k) =  \\ & \big{(} e^{\bA_{q_1}t_1} \bK_{q_{2},q_{1}} \big{)} \bg^c_{q_2,q_3,\ldots,q_k}(t_2,t_3,\ldots,t_k),
\end{align*}
it follows that the $k^{\rm th}$ Gramian corresponding to mode 1 (or respectively mode 2) can be written in terms of the $(k-1)^{\rm th}$ Gramian corresponding to mode 2 (or mode 1), as
\begin{align}\label{Pkprop}
\cP_{q_1}^{(k)} =  \int_{0}^\infty  e^{\bA_{q_1}t_1} \bK_{q_{2},q_{1}}  \cP_{q_2}^{(k-1)} \bK_{q_{2},q_{1}}^T e^{\bA_{q_1}^Tt_1} dt_1.
\end{align}

\begin{proposition}
	The level k controllability Gramians corresponding to modes 1 and 2 can be computed by iteratively solving the coupled systems of linear equations:
	\begin{align} 
	\bA_1 \cP_1^{(k)} + \cP_1^{(k)} \bA_1^T + \bK_{2,1} \cP_2^{(k-1)}  \bK_{2,1}^T  = \bfz, \label{klevel_con1} \\ \bA_2 \cP_2^{(k)} + \cP_2^{(k)} \bA_2^T  + \bK_{1,2} \cP_1^{(k-1)}  \bK_{1,2}^T = \bfz,
	\label{klevel_con2}
	\end{align}
	where $k>1$ and the starting point is represented by the linear Gramians (with no switching) $\bP_{q_1}^{(1)}$ in (\ref{lin_gram_con}) that correspond to the first level.
\end{proposition}
\noindent
{\it{Proof of Prop. 3.1}}.
By multiplying the equality in (\ref{Pkprop}) with $\bA_{q_1}$ to the left and with $\bA_{q_1}^T$ to the right, we write
$$
\bA_{q_1} \cP_{q_1}^{(k)} + \cP_{q_1}^{(k)} \bA_{q_1}^T = \int_{0}^\infty \frac{d}{dt_1} \Big{(}  e^{\bA_{q_1}t_1} \bK_{q_{2},q_{1}}  \cP_{q_2}^{(k-1)} \bK_{q_{2},q_{1}}^T 
$$
$$
e^{\bA_{q_1}^Tt_1} dt_1 \Big{)} = - \bK_{q_{2},q_{1}}  \cP_{q_2}^{(k-1)} \bK_{q_{2},q_{1}}^T.
$$
which proves the statements in (\ref{klevel_con1}) and (\ref{klevel_con2}). \sq

\subsubsection{Observability Gramians}

Introduce the level $k$ energy functional $\bg^o_{q_k,\ldots,q_2,q_1}(t_k,\ldots,t_2,t_1) : \mathbb{R}^k \rightarrow \mathbb{R}^{p_{q_k}}$, corresponding to a switching sequence  $(q_k,\ldots,q_2,q_1) \in \Omega^k$ that ends in mode $q_1$, as
\small
\begin{align}\label{goqdef}
\bg^o_{q_k,\ldots,q_1} &(t_k,\ldots,t_1) = \nonumber \\  & \bC_{q_k} e^{\bA_{q_k} t_k} \bK_{q_{k-1},q_{k}} \cdots \bK_{q_{1},q_{2}} e^{\bA_{q_1}t_1}. 
\end{align}
\normalsize
\noindent
Compute the level $k$ infinite Gramian corresponding to mode $q_1 \in \{1,2\}$  by calculating the inner product of the energy functional associated to the length $k$ switching sequence $(q_1,q_2,\ldots,q_k)$ with itself, as
\begin{align}\label{Qkdef}
\cQ_{q_1}^{(k)} &= \int_{0}^\infty \cdots \int_{0}^\infty  \big{(} \bg^o_{q_k,\ldots,q_2,q_1}(t_k,\ldots,t_2,t_1) \big{)} ^T \nonumber \\ & \bg^o_{q_k,\ldots,q_2,q_1}(t_k,\ldots,t_2,t_1) dt_1\ldots dt_k.
\end{align}

\begin{proposition}
	The level k observability Gramians corresponding to modes 1 and 2 can be computed by iteratively solving the coupled systems of linear equations (for $k>1$)
	\begin{align} 
	\bA_1^T \cQ_1^{(k)} + \cQ_1^{(k)} \bA_1 + \bK_{1,2}^T \cQ_2^{(k-1)}  \bK_{1,2}  = \bfz, \label{klevel_obs1} \\ \bA_2^T \cQ_2^{(k)} + \cQ_2^{(k)} \bA_2  + \bK_{2,1}^T \cQ_1^{(k-1)}  \bK_{2,1} = \bfz,
	\label{klevel_obs2}
	\end{align}
	where the starting point is represented by the Gramians  $\cQ_{q}^{(1)}$ in (\ref{lin_gram_obs}) that correspond to the first level (no switching).
\end{proposition}
\noindent
{\it{Proof of Prop. 3.2}}. Similar to the proof of Prop. 3.1. \sq

\subsection{The LSS case - general switching}

\begin{definition}
	Introduce the infinite controllability Gramian $\cP_{q_1}$ corresponding to mode $q_1 \in \{1,2\}$, as
	\small
	\begin{align}
	\cP_{q_1} &= \sum_{k=1}^\infty \int_{0}^\infty \cdots \int_{0}^\infty   \bg^c_{q_1,q_2,\ldots,q_k}(t_1,t_2,\ldots,t_k)  \nonumber \\ & \big{(}\bg^c_{q_1,q_2,\ldots,q_k}(t_1,t_2,\ldots,t_k) \big{)}^T  dt_1  \ldots dt_k = \sum_{k=1}^\infty \cP_{q_1}^{(k)} \label{defPgram}.
	\end{align}
	%	in terms of the multivariate functions $\bg_q^r$ in (\ref{grqdef}) or matrices $\cP_{q_1}^{(k)}$ in (\ref{Pkdef}).
\end{definition}
Note that $\cP_{q_1}$ is computed by considering the inner products of energy functionals associated to all possible switching sequences (of any length), that start in mode $q_1$.
% Moreover, by using 
%the definition in (\ref{Pkdef}), it follows that $\cP_{q_1} =\sum_{k=1}^\infty \cP_{q_1}^{(k)}$, for $q_1 \in \{1,2\}$.

\begin{proposition}
	The infinite infinite controllability Gramians defined in (\ref{defPgram}), satisfy the following system of generalizaed coupled Lyapunov equations
	\begin{equation}\label{PgramLyap}
	\begin{cases} \bA_1 \cP_1 +\cP_1 \bA_1^T+ \bK_{2,1} \cP_2 \bK_{2,1}^T+ \bB_1 \bB_1^T = \bfz, \\ \bA_2 \cP_2 +\cP_2 \bA_2^T+ \bK_{1,2} \cP_1 \bK_{1,2}^T+ \bB_2 \bB_2^T = \bfz. \end{cases}
	\end{equation}
	\vspace{0mm}
\end{proposition}
{\it{Proof of Prop. 3.3}}.
By adding the equalities stated in (\ref{klevel_con1}) and (\ref{klevel_con2}) for $k>2$ as well as the one corresponding to $k=1$ (in (\ref{def_lin_gram_con})), the results follow directly.

\begin{remark}
	Write the  matrices $\{\cP_q, \bA_q,\bB_q,\bC_q\}$, $q \in \{1,2\}$ and $\{\bK_{1,2},\bK_{2,1}\}$,  in block-diagonal format, as
	\begin{equation}\label{notationD}
	\bZ_{\bD} =\left[ \begin{array}{cc}
	\bZ_1 & \bfz \\ \bfz & \bZ_2
	\end{array} \right],  \  \bK_{\scriptsize\reflectbox{\bD}} =\left[ \begin{array}{cc}
	\bfz & \bK_{1,2}  \\  \bK_{2,1} & \bfz
	\end{array} \right],
	\end{equation} 
	where $\bZ \in \{\bA,\bB,\bC, \cP\}$. Hence, one can compactly write the two equations in (\ref{PgramLyap}) as only one equation:
	\begin{equation}\label{PgramDLyap}
	\bA_{\bD} \bP_\bD + \bP_\bD \bA_\bD^T +\bK_{\scriptsize\reflectbox{\bD}} \bP_\bD \bK_{\scriptsize\reflectbox{\bD}}^T+ \bB_\bD \bB_\bD^T = \bfz,
	\end{equation}
	and recover the controllability Gramians $\cP_1$ and $\cP_2$ as block diagonal entries of $\bP_{\bD}$.
\end{remark}

\begin{definition}
	Introduce the infinite observability Gramian $\cQ_{q_1}$ corresponding to mode $q_1 \in \{1,2\}$ of the LSS system $\Si$ as
	\small
	\begin{align}\label{defQgram}
	\cQ_{q_1} &= \sum_{k=1}^\infty \int_{0}^\infty \cdots \int_{0}^\infty \big{(}  \bg^o_{q_k,\ldots,q_2,q_1}(t_k,\ldots,t_2,t_1) \big{)}^T \nonumber \\ & \bg^o_{q_k,\ldots,q_2,q_1}(t_k,\ldots,t_2,t_1) \ dt_1  dt_2 \ldots dt_k   = \sum_{k=1}^\infty \cQ_{q_1}^{(k)} .
	\end{align}
\end{definition}
\noindent
Note that $\cQ_{q_1}$ is computed by considering the inner products of energy functionals associated to all possible switching sequences (of any length) that end in mode $q_1$.

\begin{proposition}
	The infinite observability Gramians defined in (\ref{defQgram}), satisfy the following system of generalizaed coupled Lyapunov equations
	\begin{equation}\label{QgramLyap}
	\begin{cases} \bA_1^T \cQ_1 +\cQ_1 \bA_1+ \bK_{1,2}^T \cQ_2 \bK_{1,2}+ \bC_1^T \bC_1 = \bfz \\ \bA_2^T \cQ_2 +\cQ_2 \bA_2+ \bK_{2,1}^T \cQ_1 \bK_{2,1}+ \bC_2^T \bC_2 = \bfz \end{cases}
	\end{equation}
	\vspace{0mm}
	%	in terms of the multivariate functions $\bg_q^o$ in (\ref{grqdef}) and matrices $\cQ_{q_1}^{(k)}$ in (\ref{Qkdef}).
\end{proposition}
{\it{Proof of Prop. 3.4}}. Similar to the proof of Prop. 3.3. \sq

\begin{remark}
	{\rm
		Additional to (\ref{notationD}), write the  matrices $\{\cQ_q\}, q \in \{1,2\}$  in block-diagonal format, as $\bQ_\bD =\left[ \begin{array}{cc}
		\cQ_1 & \bfz \\ \bfz & \cQ_2
		\end{array} \right]$. Again, compactly write the equations in (\ref{QgramLyap}) as only one equation:
		\begin{equation}\label{QgramDLyap}
		\bA_{\bD}^T \bQ_\bD + \bQ_\bD \bA_\bD +\bK_{\scriptsize\reflectbox{\bD}}^T \bQ_\bD \bK_{\scriptsize\reflectbox{\bD}}+ \bC_\bD^T \bC_\bD = \bfz,
		\end{equation}
		and recover the observability Gramians  as the block diagonal entries of $\bQ_{\bD}$.
	}
\end{remark}

\begin{definition}\label{def1}
	Since all eigenvalues of both $\bA_1$ and $\bA_2$ matrices have  negative real part, the same property applies for $\bA_\bD$. The system $\dot{\bx} = \bA_\bD \bx$ is asymptotically stable , or in short, $\bA_\bD$ is stable, if there exist real scalars $\beta > 0$ and $0 < \alpha  \leqslant -\max_i (\text{Re}(\lambda_i (\bA_\bD)))$, such that:
	\vspace{-2mm}
	\begin{equation*}
	\Vert e^{\bA_\bD t} \Vert \leqslant \beta e^{-\alpha t}.
	\end{equation*}
\end{definition}

\vspace{2mm}

By using the result in Theorem 2 from \cite{c16}, we address the problem of existence of the new defined Gramians. The key condition is that the norm of the coupling matrices is sufficiently small. 
\begin{proposition}
	The controllability and observability Gramians  in (\ref{defPgram}), (\ref{defQgram}) exist if
	\vspace{-2mm}
	\begin{equation}\label{existPQ}
	\begin{cases} \bA_\bD \  \text{is  stable and},  \\ \Vert \bK_\bD \Vert = \max(\Vert \bK_{1,2} \Vert, \Vert \bK_{2,1} \Vert ) \leqslant \frac{\sqrt{2 \alpha}}{\beta}.
	\end{cases}
	\end{equation}
\end{proposition}
\vspace{2mm}
Solving such generalized Lyapunov equations as (\ref{PgramDLyap}) and (\ref{QgramDLyap}) is not a straightforward task. A possible approach is to approximate these solutions with truncated sums of positive definite matrices, as
\begin{equation}
\bP_\bD \approx \sum_{k=1}^H \bP_\bD^{(k)}, \ \ \bQ_\bD \approx \sum_{k=1}^H \bQ_\bD^{(k)}, \ \ H \geqslant 1,
\end{equation}
where one can find $\bP_\bD^{(k)}$ by solving the Lyapunov equation $\bA_{\bD} \bP_\bD^{(k)} + \bP_\bD^{(k)} \bA_\bD +\bK_{\scriptsize\reflectbox{\bD}} \bP_\bD^{(k-1)} \bK_{\scriptsize\reflectbox{\bD}} ^T = \bfz$, $k \geqslant 2$. Similar equations can be solved to find $\bQ_\bD^{(k)}$.

\section{The $\mathcal{H}_2$ norm for LSS}

\subsection{The case with no switching}

Assume as before, that the system $\Si$ operates only in mode $q \in \{1,2\}$. It follows that the  $\mathcal{H}_2$ norm of the linear subsystem $\Si_q$ can be defined in time domain, in terms of the impulse response $\bh_q(t) = \bC_q e^{\bA_q t} \bB_q$, as:
\begin{align}\label{H2norm_lin_def}
\Vert \Si_q \Vert_{\mathcal{H}_2}^2 = \tr \Big{(} \int_0^\infty \bh_q(t) \bh_q^T(t) dt \Big{)}.
\end{align}
Note that one can show that the above introduced quantity can be written in terms of the Gramians defined in (\ref{def_lin_gram_con}) and (\ref{def_lin_gram_obs}), as follows:
\begin{align}\label{H2norm_lin_prop}
\Vert \Si_q \Vert_{\mathcal{H}_2}^2 = \tr(\bC_q \cP_q \bC_q^T) = \tr(\bB_q^T \cQ_q \bB_q).
\end{align}

\subsection{Extension to the class of LSS}

We propose an extension of the $\mathcal{H}_2$ norm definition in (\ref{H2norm_lin_def}) for linear switched systems, by taking into consideration all possible switching scenarios.
\begin{definition}
	Let $\Si$ be a LSS with $M=2$ modes, as introduced in (\ref{LSS_def0}) . Consider $\delta = (q_1,q_2,\ldots,q_k) \in \Omega^k$ to be a switching sequence of length $k \geqslant 1$ and  the generalized kernel functions $\bh_\delta:\mathbb{R}^k \rightarrow \mathbb{R}^p$ in (\ref{kern}) corresponding to the sequence $\delta \in \Omega^k$. By computing the inner products summation of such kernels  corresponding to all switching sequences, we define the following norm:  
	\small
	\begin{align}\label{H2norm_lss_def}
	\hspace{-2mm} \Vert \Si \Vert_{\mathcal{H}_2}^2 &= \tr \Big{(} \sum_{\delta \in \Omega^k}\int_0^\infty \cdots \int_0^\infty \bh_{q_1,q_2,\ldots,q_k}(t_1,t_2,\ldots,t_k) \nonumber \\ &
	\big{(} \bh_{q_1,q_2,\ldots,q_k}(t_1,t_2,\ldots,t_k) \big{)} ^T dt_1 dt_2 \ldots dt_k \Big{)}.
	\end{align}
	\normalsize
\end{definition}
The explicit computation of the quantity proposed in (\ref{H2norm_lss_def}) seems to be a very tedious task. To address this issue, we propose an extension of the results in (\ref{H2norm_lin_prop}). The following result states that the $\cH_2$ norm  can be expressed in terms of the
Gramians $\cP_q$ and $\cQ_q$, provided that these matrices exist. In this way, computing the proposed norm is conditioned by solving linear matrix equations.

\begin{proposition}
	Let $\Si$  be a LSS with 2 modes as in (\ref{LSS_def0}). Let $\cP_q$ be the controllability Gramians defined as in (\ref{defPgram})  which satisfy the equations in (\ref{PgramLyap}). Then we can rewrite the $\cH_2$ norm of the system $\Si$ as follows:
	\begin{equation}\label{normPgram}
	\Vert \Si \Vert_{\mathcal{H}_2}^2 = \tr(\bC_1 \cP_1 \bC_1^T) + \tr(\bC_2 \cP_2 \bC_2^T).
	\end{equation}
	Additionally, consider $\cQ_q$ be the observability Gramians defined  in (\ref{defQgram}) that satisfy the equations in (\ref{QgramLyap}). Then the following also holds:
	\begin{equation}\label{normQgram}
	\Vert \Si \Vert_{\mathcal{H}_2}^2 = \tr(\bB_1^T \cQ_1 \bB_1) + \tr(\bB_2^T \cQ_2 \bB_2).
	\end{equation}
	\vspace{-4mm}
\end{proposition}
{\it{Proof of Prop. 4.1}}. Split the right hand side of the equality in (\ref{H2norm_lss_def}) in two categories, corresponding to sequences $\delta$ that start in mode 1 and respectively, in mode 2. Hence write:
\small
\begin{align*}
\Vert \Si \Vert_{\mathcal{H}_2}^2 = \tr \Big{(} \sum_{\delta \in \Omega^k, \delta_1 = 1}\int_0^\infty  \bh_{\delta} \bh_{\delta}^T + \sum_{\delta \in \Omega^k, \delta_1 = 2}\int_0^\infty  \bh_{\delta} \bh_{\delta}^T \Big{)}
\end{align*}
\normalsize
which together with the definition of $\cP_q$ in (\ref{defPgram}), directly proofs the result in (\ref{normPgram}). Similarly, by again splitting the right hand side of (\ref{H2norm_lss_def}) in two categories, corresponding to sequences $\delta$ that end in mode 1 and respectively, in mode 2, one can also proof the result in (\ref{normQgram}). \sq

\begin{corollary}
	Additionally, by making use of the block diagonal matrices $\cP_\bD$ and $\cQ_\bD$, one can write the $\cH_2$ norm of the system $\Si$, in the following way:
	\begin{equation}
	\Vert \Si \Vert_{\mathcal{H}_2}^2 = \tr \big{(} \bC_\bD \cP_\bD \bC_\bD^T \big{)} =  \tr \big{(} \bB_\bD^T \cQ_\bD \bB_\bD \big{)}.
	\end{equation}
\end{corollary}
\vspace{2mm}
This result directly follows from (\ref{normPgram}) and (\ref{normQgram}) by taking into account the special structure of the matrices in (\ref{notationD}).

\section{The proposed method and its application}

Linear switched systems have common features with bilinear systems. For example, the LSS kernels in (\ref{kern}) have similar format to the bilinear kernels in \cite{c16} (Section 3, (14)). 
%Additionally, the generalized transfer functions of the LSS are similar to the bilinear ones.
Moreover, given that $n_1=n_2$, $m_1=m_2$ and all coupling matrices are identity matrices, it is possible to formulate an LSS as a bilinear system with fixed inputs such as:
% in multiple input format (with additional fixed inputs that alternate between 0 and 1 to activate or deactivate the $\bA_k$ matrices).
% Introduce the signal $\hat{\bu}(t) = q-1, \ $ when mode $q \in \{1,2\}$ is activated. 
\begin{align*}
\hat{\bu}(t)  = \begin{cases}
0, \ \ t \in (T_k,T_{k+1}] \ \text{so that} \ \sigma(t) = 1, \\  1, \ \ t \in (T_k,T_{k+1}] \ \text{so that} \ \sigma(t) = 2.
\end{cases}
\end{align*}
%\begin{align}
%\dot{\bx}(t) = \bA_1 \bx(t) + & (\bA_2 - \bA_1)  \bx(t) \hat{\bu}(t) \bB_1\bu(t) (1- \hat{\bu}(t))+ \bB_2 \bu(t) \hat{\bu}(t)
%\end{align}
%or equivalently, to emphasize the bilinear multiple input format, as
Rewrite the dynamics of $\Si$, as introduced in (\ref{LSS_def0}), as follows:
\begin{align}
&\dot{\bx}(t) = \bA^{\text{bil}} \bx(t)+ \bN_1^{\text{bil}}  \bx(t) \bu(t) + \bN_2^{\text{bil}}  \bx(t) \hat{\bu}(t) \label{eqbil}  \\ &+ \bN_3^{\text{bil}}  \bx(t) \bu(t) \hat{\bu}(t) + \bB_1^{\text{bil}} \bu(t) + \bB_2^{\text{bil}} \hat{\bu}(t) + \bB_3^{\text{bil}} \bu(t) \hat{\bu}(t), \nonumber
\end{align}
or equivalently, with the specific bilinear structure as,
\begin{align*}
\dot{\bx}(t) = \bA^{\text{bil}} \bx(t) + \sum_{i=1}^{3} \bN_i^{\text{bil}} \bx(t) \bu_i(t) + \sum_{i=1}^{3} \bB_i^{\text{bil}}  \bu_i(t),
\end{align*}
where $ \bA^{\text{bil}} = \bA_1, \ \bN_2^{\text{bil}} = \bA_2 - \bA_1, \ \bN_1^{\text{bil}} = \bN_3^{\text{bil}} = \bfz, \ \bB_1^{\text{bil}} = \bB_1, \ \bB_2^{\text{bil}} = \bfz, \  \bB_3^{\text{bil}} = \bB_2- \bB_1$,
%Note that the equation in (\ref{eqbil}) has bilinear structure, and can be written as
and the 3 control inputs are $\bu_1 = \bu$, $\bu_2 = \hat{\bu}$ and $\bu_3 = \bu \hat{\bu}$. 

\subsection{An iterative algorithm for LSS MOR}

Based on the above mentioned similarities, our proposed algorithm, i.e Algorithm 1 (entitled Sw-IRKA), can be viewed as a direct extension of Algorithm 2 in \cite{c5} (entitled B-IRKA), to the class of LSS.

In Algorithm 1, the input arguments are given by the system matrices of the original LSS in (\ref{LSS_def0}). Additionally, we require the  desired reduced order $r_j$ corresponding to  subsystem $\Si_j$ for $j\in \{1,2\}$. Furthermore, we initialize the algorithm with random matrices $\hat{\bA}_j \in \mathbb{R}^{r_j \times r_j}, \hat{\bB}_j \in \mathbb{R}^{r_j \times m_j}, \hat{\bC}_j \in \mathbb{R}^{p_j \times r_j}$, and $\hat{\bK}_{1,2}, \hat{\bK}_{2,1}^T \in   \mathbb{R}^{r_2 \times r_1}$. 

We use the same notations as in (\ref{notationD}) for the block diagonal matrices used in Algorithm 1. First, diagonalize the matrix $\hat{\bA}_\bD$, and let $\Lambda_\bD \in \mathbb{R}^{(n_1+n_2)\times(n_1+n_2)}$ be the diagonal eigenvalue matrix. Transform the other matrices by using the eigenvector matrix $\bU_\bD$. Next, compute projection matrices $\bX_\bD, \bX_\bD \in \mathbb{R}^{(n_1+n_2)\times(r_1+r_2)} $ by solving two generalized Lyapunov equations (see steps 4 and 5). Afterwards, modify these projectors by constructing an orthonormal basis for the range of both $\bX_\bD$ and  $\bX_\bD$ followed by imposing the condition $\bY_{\bD}^T \bX_{\bD} = \bI_{r_1+r_2}$. Finally, at step 7 we project with new matrices constructed at step 6 and come up with a reduced order LSS model $\hat{\Si}$ described by:
\begin{align*}
\hat{\Si} = (r_1,r_2, \{(\hat{\bA}_q, \hat{\bB}_q, \hat{\bC}_q)| q \in \Omega\},\{\hat{\bK}_{1,2}, \hat{\bK}_{2,1} \}, \bfz).
\end{align*}
Repeat this procedure until the eigenvalues of the matrix $\hat{\bA}_\bD$ are constant (the deviation with respect with the previous step does not exceed a certain tolerance value $\epsilon >0$).

\begin{algorithm}[H]
	\footnotesize
	\caption{\small{IRKA type approach for LSS: Sw-IRKA}}
	\begin{algorithmic}[1]
		\Procedure{}{}\newline
		\textbf{Input: $\bA_j, \bB_j, \bC_j, \bK_{i,j}, \hat{\bA}_j, \hat{\bB}_j, \hat{\bC}_j, \hat{\bK}_{i,j}, r_j, i, j \in \{1,2\}$.} \newline
		\textbf{Output: $\hat{\bA}_j^{\rm end}, \hat{\bB}_j^{\rm end}, \hat{\bC}_j^{\rm end}, \hat{\bK}_j^{\rm end}, \ j \in \{1,2\}$.}
		%       \State{Solve $\begin{cases}  \bA_1 \bP_1+\bP_1 \bA_1^T+ \bK_2 \bP_2 \bK_2^T+\bB_1 \bB_1^T = \mathbf{0} \\  \bA_2 \bP_2+\bP_2 \bA_2^T+\bK_1 \bP_1 \bK_1^T+\bB_2 \bB_2^T = \mathbf{0} \end{cases}$. }
		
		\State {while (change in $\mathbf{\Lambda_\bD} > \epsilon)$ \ do .}
		
		\State {Find $\hat{\bA}_\bD = \bU_\bD \mathbf{\Lambda}_\bD \bU_\bD^{-1}, \ \tilde{\bA}_\bD := \mathbf{\Lambda}_\bD,  \  \tilde{\bB}_\bD = \bU_\bD^{-1} \hat{\bB}_\bD$, $\tilde{\bC}_\bD =  \hat{\bC}_\bD \bU_\bD, \ \ \tilde{\bK}_{\scriptsize\reflectbox{\bD}} =  \bU_\bD^{-1} \hat{\bK}_{\scriptsize\reflectbox{\bD}} \bU_\bD$.} 
		
		\State {Solve $\bA_\bD \bX_\bD+\bX_\bD \tilde{\bA}_\bD^T+\bK_{\scriptsize\reflectbox{\bD}} \bX_\bD \tilde{\bK}_{\scriptsize\reflectbox{\bD}}^T+ \bB_\bD \tilde{\bB}_\bD^T = \bfz$.}

		\State {Solve $\bA_\bD^T \bY_\bD+\bY_\bD \tilde{\bA}_\bD+\bK_{\scriptsize\reflectbox{\bD}}^T \bY_\bD \tilde{\bK}_{\scriptsize\reflectbox{\bD}}+ \bC_\bD^T \tilde{\bC}_\bD = \bfz$.}

		\State {$\bX_\bD = \text{\rm orth}(\bX_\bD), \bY_\bD = \text{\rm orth}(\bY_\bD), \bY_\bD = \bY_\bD (\bX_\bD^T \bY_\bD)^{-1} $.}
		
		\State {$\hat{\bA}_\bD =  \bY_\bD^T \bA_\bD \bX_\bD, \ \ \hat{\bB}_\bD = \bY_\bD^T \bB_\bD, \ \ \hat{\bC}_\bD = \bC_\bD \bX_\bD, \ \ \hat{\bK}_{\scriptsize\reflectbox{\bD}} =  \bY_\bD^T \bK_{\scriptsize\reflectbox{\bD}} \bX_\bD$.}
		
		%		\State {$\hat{\bA}_\bD = (\bY_\bD^T \bX_\bD)^{-1} \bY_\bD^T \bA_\bD \bX_\bD, \ \ \hat{\bB}_\bD = (\bY_\bD^T \bX_\bD)^{-1} \bY_\bD^T \bB_\bD, \ \ \hat{\bC}_\bD = \bC_\bD \bX_\bD, \ \ \hat{\bK}_{\scriptsize\reflectbox{\bD}} = (\bY_\bD^T \bX_\bD)^{-1} \bY_\bD^T \bK_{\scriptsize\reflectbox{\bD}} \bX_\bD$.}

		\State {end while.}
		
		\State {Return  $\hat{\bA}_\bD^{\rm end} = \hat{\bA}_\bD,  \ \hat{\bB}_\bD^{\rm end} = \hat{\bB}_\bD,  \ \hat{\bC}_\bD^{\rm end} = \hat{\bC}_\bD,  \ \hat{\bK}_{\scriptsize\reflectbox{\bD}}^{\rm end} = \hat{\bK}_{\scriptsize\reflectbox{\bD}} $.}        
		
		\EndProcedure
	\end{algorithmic}
\end{algorithm}

\normalsize

\subsection{A numerical example}

Consider the CD player system from the benchmark examples for MOR in \cite{c8}, which is a linear system of order 120 with two inputs and two outputs. Assume that, at any given instance of time, only one input and one output are active (the others are not functional due to, for example, a mechanical failure). More precisely, consider mode $j$ to be activated whenever the $j^{th}$ input and the $j^{th}$ output are simultaneously failing (where $j \in \{1,2\})$. Hence, construct a LSS system $\Si$ with two operational modes and stable subsystems of order $n_1 = n_2 = 120$. The impulse response of each original subsystem $\Si_j$ is depicted in Fig.\;1. 

\begin{figure}[thpb]
	\centering
	\framebox{\parbox{3in}{\includegraphics[scale=0.22]{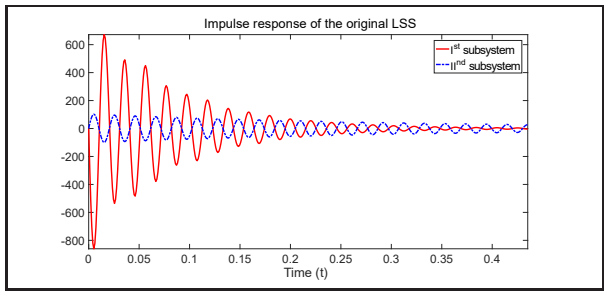}
	}}
	\caption{Impulse response of the two subsystems}
	\label{fig1}
\end{figure}

The system $\Si$ is reduced using the method described in Algorithm 1 to obtain $\hat{\Si}_{\rm IR}$. Additionally, compute a reduced order model $\hat{\Si}_{\rm BT}$ by means of the balanced truncation (BT) method proposed in \cite{c12}. The orders of the reduced subsystems are set to be $r_1 = r_2 = 18$ for both reduced models. In the following numerical experiment, the continuous control input is chosen to be $\bu(t) = (1 + \sin(\pi t)) e^{-t/5}, t \geqslant 0$. By considering a switching sequence that starts in mode $q_1 = 1$ with randomly chosen switching times $T_j$ in the time interval $[0,10]s$, we perform a time domain simulation. In Fig.\;2,  the switching signal $\sigma(t)$ is depicted in the upper part, while in the lower part of the figure, the outputs corresponding to the original and the reduced LSS are presented.

\begin{figure}[thpb]
	\centering
	\framebox{\parbox{3in}{\includegraphics[scale=0.22]{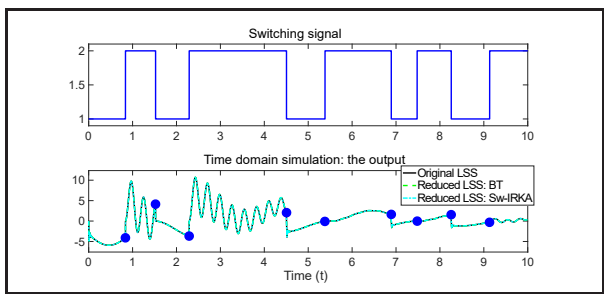}
	}}
	\caption{The switching signal and the observed output}
	\label{fig2}
\end{figure}

Furthermore, the approximation error of the original output for both MOR methods is depicted in Fig.\;3. We observe that the error corresponding to the new proposed method is in general lower than the one produced by the BT method (for most of the time instances included in the simulation).

\begin{figure}[thpb]
	\centering
	\framebox{\parbox{3in}{\includegraphics[scale=0.22]{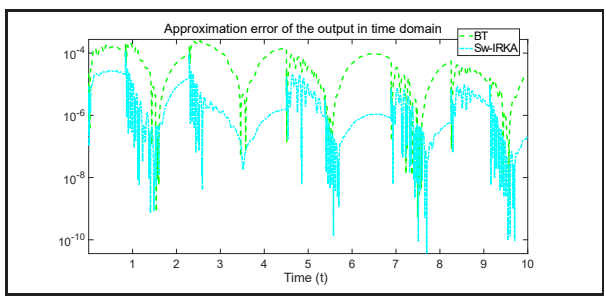}
	}}
	\caption{Time domain approximation error}
	\label{fig3}
\end{figure}

As described in Section 5.1, the stopping criterion of the Sw-IRKA method is that the poles of the reduced model are essentially the same (up to rounding errors quantified by a tolerance $\epsilon$).
Since the proposed method is iterative, we present an analysis of the number of iterations needed to reach an offset of $\epsilon = 10^{-8}$. As it can be observed in Fig.\;4, 35 iterations were sufficient for both subsystems.

\begin{figure}[thpb]
	\centering
	\framebox{\parbox{3in}{\includegraphics[scale=0.22]{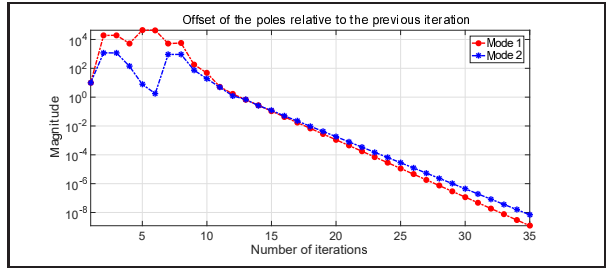}
	}}
	\caption{The pole offset after each iteration for both modes}
	\label{fig5}
\end{figure}
Finally, we choose the same truncation orders for the two subsystems, i.e. $r_1 = r_2 = r$. This value is varied in steps of 5 in between the range [5,30]. For each $r$, the relative $\cH_2$ error is computed for both analyzed methods. The results are collected in Table 1. As expected, the new proposed method produces lower errors in all the cases. 

\begin{table}[h] 
	\caption{Relative $\cH_2$ error of the two reduced LSS}
	\vspace{-4mm}
	\label{table_example}
	\begin{center}
		%\small
		\begin{tabular}{|c|Sc|Sc|}
			%{|M{2.5cm}|M{3.2cm}|M{3.2cm}|}
			\hline
			r  & $\Vert \Si_{\rm BT} - \Si \Vert_{\mathcal{H}_2} / \Vert \Si \Vert_{\mathcal{H}_2} $ & $\Vert \Si_{\rm IR} - \Si \Vert_{\mathcal{H}_2} / \Vert \Si \Vert_{\mathcal{H}_2} $  \\
			\hline 
			$5$ &  $6.77 \cdot 10^{-3}$   & $1.62 \cdot 10^{-3}$ \\
			\hline 
			$10$  & $1.83 \cdot 10^{-4}$    & $7.12 \cdot 10^{-5}$ \\
			\hline
			$15$ &  $5.87 \cdot 10^{-5}$ &  $2.46 \cdot 10^{-5}$  \\
			\hline
			$25$ &  $4.95 \cdot 10^{-5}$   & $8.96 \cdot 10^{-6}$  \\
			\hline
			$30$ &  $1.64 \cdot 10^{-5}$   & $4.74 \cdot 10^{-6}$  \\
			\hline
		\end{tabular}
	\end{center}
\end{table}

\section{Conclusions and further developments}

In this work, a definition of the $\cH_2$ norm was proposed for the class of linear switched systems. Although the special case of systems with two modes was considered, the results can easily be extended to the general case of $M$ modes. Moreover, an extension of the reduction method known in the literature as IRKA, was introduced in Algorithm 1. The results provided in Section 5.2 show that our method could be successfully applied for a benchmark example. The approximation quality turned out to be in general better than that of the balanced truncation method in \cite{c12}.

Finding appropriate optimality conditions for the proposed LSS reduction framework  (as the ones found for the bilinear systems case, i.e. (4.5)-(4.8) in \cite{c5} as well as (4.12)-(4.13) in \cite{c9}) can be a possibly reasonable topic of further research.


%%%%%%%%%%%%%%%%%%%%%%%%%%%%%%%%%%%%%%%%%%%%%%%%%%%%%%%%%%%%%%%%%%%%%%%

\end{document}